\documentclass[conference,compsoc]{IEEEtran}
\usepackage{cite}
\usepackage{amsmath,amssymb,amsfonts}
\usepackage{algorithmic}
\usepackage{graphicx}
\usepackage{textcomp}
\usepackage{xcolor}
\usepackage{tikz}
\usepackage{booktabs}
\usepackage{url}
\usepackage{tabularx}
\usepackage{stfloats}
\usepackage{microtype}

\usetikzlibrary{positioning, arrows.meta, fit, calc}
\def\BibTeX{{\rm B\kern-.05em{\sc i\kern-.025em b}\kern-.08em
    T\kern-.1667em\lower.7ex\hbox{E}\kern-.125emX}}
\begin{document}

\title{Closing the Sim-to-Real Gap: An Evaluation Framework for Autonomous Cyber Defense Configuration of Commercial EDR}
\author{\IEEEauthorblockN{Kerri Prinos$^{*}$}
\IEEEauthorblockA{\textit{Horizon3} \\
San Francisco, CA, USA \\
\ kerri.prinos@Horizon3.ai
}
\and
\IEEEauthorblockN{Lilianne Brush$^{*}$}
\IEEEauthorblockA{\textit{Horizon3} \\
San Francisco, CA, USA \\
\ lili.brush@Horizon3.ai}
}

\maketitle
\def\thefootnote{*}\footnotetext{These authors contributed equally to this work}\def\thefootnote{\arabic{footnote}}
\begin{abstract}Leading commercial endpoint detection and response (EDR) products have shifted from operator-configured rule sets to multi-component systems where autonomous AI components operate alongside, and increasingly in place of, operator-deployed policies. Autonomous defense agents using commercial EDR as their hardening tool are no longer tuning a passive tool, but a black-box autonomous system capable of making vendor-specific decisions. We present the first evaluation framework for autonomous defense agents hardening commercial EDR. We instantiate it in a Game of Active Directory (GOAD) lab with Horizon3's NodeZero as the autonomous pentester and Microsoft Defender XDR as the primary EDR under test. We extend the same framework to CrowdStrike Falcon and SentinelOne Singularity to test whether our Microsoft Defender XDR findings hold across vendors. We run a sample benchmark of defense agents with two large language model (LLM) backbones (Claude Sonnet 4.6 and Cisco Foundation-Sec-8B). We report three lessons learned that neither simulation nor open source EDR evaluation can surface: (i) commercial EDR telemetry is engineered for Security Operations Center (SOC) analyst workflows rather than for measuring policy effectiveness; (ii) per-policy attribution is important for separating the defense agent's actions from the EDR's autonomous actions; and (iii) EDR behavior varies between runs, largely because containment varies and accumulates. Together, these findings highlight a sim-to-real gap for enterprise defense and motivate evaluation methodology for benchmarking autonomous defense agents in environments with black-box, autonomous tools.
\end{abstract}


\section{Introduction}\label{sec:intro}
The capabilities of large language model (LLM)-driven offensive cyber agents are rapidly advancing, and commercial endpoint detection and response (EDR) products are a primary defense against them. The 2025 Gartner Magic Quadrant for Endpoint Protection Platform leaders, Microsoft, CrowdStrike, SentinelOne, Palo Alto, Trend Micro, and Sophos~\cite{gartner}, together account for 57.7\% of the enterprise market share~\cite{IDC}. Each of these products exposes dozens to hundreds of configurable detection and blocking policies. Literature on EDR performance benchmarking and policy optimization relies on vendor-reported policy effectiveness, open-source EDR, or undisclosed defensive configurations.

Each new evaluation framework increases in fidelity to real-world environments. The progression runs from pure simulation and emulation~\cite{CybORG, Cage4} to emulation with industry-level tools~\cite{janisch}, and most recently to autonomous cyber defense agents with open-source EDR to harden their environments~\cite{dreadGOAD, mayoral-vilches}. However, there is still a critical sim-to-real gap between existing evaluation frameworks and the complexities of modern enterprise environments. Leading commercial EDR products, such as Microsoft Defender XDR~\cite{ms-defender-xdr-product}, CrowdStrike Falcon~\cite{cs-falcon}, and SentinelOne Singularity~\cite{s1-singularity}, have shifted from operator-configured rule sets to multi-component systems in which vendor-managed autonomous AI components operate alongside (or in the place of) operator-deployed policies. The most advanced cyber defense evaluation frameworks~\cite{dreadGOAD},~\cite{mayoral-vilches} use open-source EDR, which is rule-based and operator-driven and does not surface any of these autonomous-component dynamics. For modern enterprise defense, we have to consider that cyber defense agents are now interacting with and alongside other autonomous, vendor-specific, black-box components.

We present an autonomous, end-to-end evaluation framework for cyber defense agents that orchestrates closed-loop experiments with integrated commercial EDR. The experiment controller executes four sequential stages: Preflight, Setup, Loop, and Teardown. A model-agnostic proposer wraps the defense agent being evaluated, so our system could be used to benchmark performance across different agentic architectures and models.

For this initial implementation, we use a Game of Active Directory (GOAD) lab~\cite{goad} as the environment, Horizon3's NodeZero~\cite{nodezero} as the autonomous pentester, and Microsoft Defender XDR~\cite{ms-defender-xdr-product} as the EDR. We run a sample benchmark of two LLM-based defense agents. To test whether our findings generalize beyond one vendor, we also run the framework on CrowdStrike Falcon and SentinelOne Singularity. The lab, pentester, and defense agents stay the same; only the vendor-specific correlation methods and policy options change. Our core contributions are lessons learned and recommendations for evaluating cyber defense agents in enterprise environments. We use the benchmark results to illustrate commercial EDR behaviors that simulations and open-source EDR do not capture.

\section{Background}\label{sec:background}
In this section, we motivate the need for an evaluation framework for autonomous defense agents that configure commercial EDR. After detailing EDR fundamentals, we examine gaps in MITRE ATT\&CK~\cite{mitre_attack} technique coverage, differences in vendor-claimed and observed EDR behavior, false positives and lack of per-policy attribution within EDR output, and black-box, vendor-managed autonomous features.

\subsection{Commercial EDR Fundamentals} Endpoint Detection and Response (EDR) software runs a sensor on each host. The sensor watches process, file, registry, and network activity and reports it to a vendor cloud as telemetry. The cloud analyzes this telemetry and raises alerts on malicious behavior~\cite{karantzas}. The sensor can also block actions outright. Operators tune what the EDR blocks through policies: named rules aimed at attacker techniques, such as stealing credentials from memory. Extended Detection and Response (XDR) adds identity, email, and cloud signals to the same console. Modern products also run autonomous components without an operator rule. For example, Microsoft Defender XDR's Automated Investigation and Response (AIR)~\cite{microsoft_air} investigates and remediates alerts, and its Automatic Attack Disruption~\cite{ms-attack-disruption} contains compromised assets (isolating hosts, containing user accounts). The containment we observe (\texttt{ContainedUser*}) comes from the latter. These alerts reach a Security Operations Center (SOC) analyst, who reads them to investigate incidents. In this work, an autonomous defense agent takes the operator's role: it reads the attack outcome and deploys policies before the next attack. We collect the acronyms and domain terms used throughout in the Glossary (Appendix~\ref{app:glossary}).

\subsection{Attack Coverage Gap} Recent research shows that commercial EDRs have coverage gaps for both individual MITRE ATT\&CK techniques and multi-step, chained attacks. Because EDRs do not natively reason about kill-chain causality, they fragment multi-step attacks into single-step alerts~\cite{shen,hassan}. Analyzing the MITRE Engenuity Enterprise Evaluations across 37 vendors and 16.4k detection results, Shen et al.~\cite{shen} conclude that single-step alerts do not give EDRs enough confidence to act, and that attack-graph-level correlation is necessary for effective response. Virkud et al.~\cite{virkud} found in 2024 that the Carbon Black, Splunk, and Elastic rule sets cover only about 48--55\% of MITRE ATT\&CK techniques, and that covering a technique does not guarantee catching the real-world attacks that use it.

\subsection{Expected vs. Observed Behavior Gap} Vendor-claimed EDR effectiveness diverges substantially from observed performance under real-world attacks. Karantzas and Patsakis~\cite{karantzas} empirically assessed eleven commercial EDR products against four Advanced Persistent Threat (APT) scenarios and found that most neither block nor log the majority of attacks. They call for comprehensive event analysis, and flag the lack of systematic measurement of EDR true positives, false negatives, false positives, and response time under real-world attacks as a critical gap in the literature. The EDR Telemetry Project~\cite{telemetry_project} increases transparency by running controlled actions and recording raw telemetry, but it tests only default, out-of-the-box configurations and does not evaluate policy tuning.

\subsection{False Positives and Per-Policy Attribution Gap} In a four-year longitudinal study of real-world network alert logs from SOCs, Yang et al.~\cite{yang} found that analysts are overwhelmed with 24k to 134k alerts per day, but only 0.01\% of the alerts are associated with true attacks. In the workforce, more than 25\% of analysts' time is spent handling false positives~\cite{tariq}, contributing to alert fatigue and burnout~\cite{tariq},~\cite{alahmadi}. The EDR Telemetry Project~\cite{telemetry_project} records which techniques each policy covers, but not how much each policy adds to the false-positive load.

\subsection{Configuration Space and Tuning Cost}
EDR policy optimization is a combinatorially intractable problem. Consider Microsoft Defender for Endpoint (MDE), which has more than 100 baseline policy settings, each with multiple options~\cite{intune_settings}. For simplicity, assume each setting has just two modes, enabled and disabled. Even then, the baseline settings alone yield more than $2^{100}$ or $1.26\times10^{30}$ possible configurations before per-host filtering, interaction effects, or non-baseline variations. Microsoft's deployment guidance recommends starting every policy in audit mode, observing false-positive volume over a 30-day baseline, and adding per-environment exclusions for line-of-business applications before promoting to block~\cite{microsoft_recs}. This 30-day audit-then-enforce cadence functions as an effective maintenance-window budget: an enterprise can responsibly introduce only a limited number of policy updates per cycle without risking line-of-business disruption~\cite{microsoft_asr_faq}. Because tuning depends on each enterprise's environment and business needs, no single configuration is optimal for all of them.

\subsection{Black Box Autonomous Features}
Today's leading commercial EDR and XDR products, such as Microsoft Defender XDR, CrowdStrike Falcon, and SentinelOne Singularity, embed autonomous detection and response capabilities that operate alongside or in place of operator-configured rule-based policies. The granularity of operator control over these autonomous layers varies by vendor. For example, Microsoft Defender XDR's AIR lets operators set only how much it automates (full automation by default~\cite{microsoft_air}); its automatic attack disruption is tuned only by which users, devices, and IPs to exclude. Operators cannot see why attack disruption contains a given user or host: the classifier behind its \texttt{ContainedUser*} actions is not exposed. Across vendors, the decision logic is proprietary and opaque, and the operator's controls are tenant-level toggles. An agent could in principle treat these toggles as part of its action space. Doing so requires both a vendor API that exposes them and a methodology that attributes outcomes to the agent's policy choices separately from the EDR's own AI-driven behavior. Across all three vendors, this paper restricts the agent's action space to operator-deployed rule-based policies; on Microsoft Defender XDR, it also leaves AIR at its default full-automation setting.

\begin{figure}[!t]
\centering
\resizebox{\columnwidth}{!}{%
\begin{tikzpicture}[
    node distance=1.4cm and 2.2cm,
    box/.style={draw, rounded corners, minimum width=2.4cm, minimum height=0.9cm, align=center, font=\small},
    component/.style={box},
    controller/.style={box, minimum width=2.8cm},
    arr/.style={-{Stealth[length=2.5mm]}, thick},
    biarr/.style={{Stealth[length=2.5mm]}-{Stealth[length=2.5mm]}, thick},
    lbl/.style={font=\scriptsize, midway, fill=white, inner sep=1pt},
]
\node[controller] (ctrl) {Experiment Controller};

\node[component, left=of ctrl, yshift=1.2cm] (goad) {Environment: GOAD Lab\\{\scriptsize 8 hosts}};
\node[component, left=of ctrl, yshift=-1.2cm] (n0) {Attacker: NodeZero};

\node[component, right=of ctrl, yshift=1.2cm] (defender) {EDR/XDR: Microsoft Defender XDR};
\node[component, right=of ctrl, yshift=-1.2cm] (redshift) {Data Store: Redshift};

\node[component, below=1cm of ctrl, minimum width=3.2cm] (proposer) {Proposer (Variable)};

\draw[arr] (ctrl) -- node[lbl, above, sloped] {WinRM/SSH} (goad);
\draw[arr] (ctrl) -- node[lbl, below, sloped] {H3 CLI} (n0);
\draw[arr] (n0.south) |- ++(0,-1.2) -| node[lbl, below, pos=0.25] {attack data} (redshift.south);
\draw[arr] (defender) -- node[lbl, above, sloped] {KQL} (ctrl);
\draw[arr] (redshift) -- node[lbl, below, sloped] {telemetry} (ctrl);
\draw[biarr] (ctrl) -- node[lbl, right] {propose / picks} (proposer);
\end{tikzpicture}}
\caption{System architecture. Arrows show data flow. The controller deploys policies to the GOAD hosts and launches NodeZero pentests against them. Offensive telemetry flows from NodeZero to Redshift, and defensive telemetry flows from Microsoft Defender XDR to the controller. The proposer wraps the defense agent under test.}
\label{fig:architecture}
\end{figure}

\section{Related Work}
Cyber evaluation frameworks have evolved from basic simulation and emulation to complex, dynamic environments with multi-agent support and real-world tool integration. The CybORG~\cite{CybORG} and the CAGE Challenge series~\cite{Cage4} initially adapted the observation/action/reset/step interface from OpenAI Gym~\cite{openai_gym}. These cyber gyms support algorithm development and train red- and blue-team reinforcement learning (RL) agents in simulated and emulated environments. NASimEMu~\cite{janisch} added emulation tools (Vagrant, VirtualBox, and Metasploit) so that deep RL cyber agents trained in simulation transfer better to real systems. Shifting from training to evaluation, the MITRE ATT\&CK Engenuity Evaluations benchmark commercial EDRs against emulated APT attacks, reporting detection and protection quality for each vendor~\cite{engenuity}. However, the public results do not disclose the specific policy changes vendors make to improve their scores~\cite{engenuity,shen}. Outkin et al.~\cite{outkin} provide one of the first realistic offline evaluations of defender behavior. They ground it in attack-success metrics and defender-response data from the Engenuity Evaluations, building a game-theoretic model to optimize defense policy and resource allocation.

The closest prior art deploys autonomous attackers against defender agents that harden their environment by tuning open-source EDR. Dreadnode released an open-source, closed-loop evaluation framework combining DreadGOAD, a vulnerable Active Directory lab, and Ares, an autonomous multi-agent system, to measure the performance of red and blue team agents~\cite{dreadGOAD}. DreadGOAD~\cite{dreadGOAD_repo} includes plug-in extensions for open-source EDR (Wazuh). Mayoral-Vilches et al. introduced Dynamic Cyber Ranges to evaluate LLM-driven, APT attackers while LLM-driven defender agents harden the environment using open-source security tools~\cite{mayoral-vilches}. Open-source EDRs, like Wazuh~\cite{wazuh} and Velociraptor~\cite{velociraptor}, are rule-based and operator-driven, with no autonomous-response layer, so their policy tuning differs fundamentally from that of commercial products. Existing cyber gyms and frameworks built on open-source EDR therefore cannot capture the challenges of evaluating commercial-EDR configurations. Gartner's commercial EDR leaders account for 57.7\% of the enterprise market~\cite{IDC}. In this work, we bring agent evaluation closer to real enterprise practice with the first evaluation framework for cyber defense agents that configure commercial EDR.

\section{Approach}\label{sec:methods}

\subsection{System Overview}\label{sec:system-overview}

We present a closed-loop evaluation framework for autonomous tuning of commercial EDR. A proposer (e.g., an LLM-based defense agent) selects policy configurations; the framework deploys them, measures each change by running a commercial penetration test (pentest), and feeds the resulting telemetry back to the proposer for the next round. We designed an open-architecture framework, and our first implementation uses a fixed environment, data store, attacker, and commercial EDR/XDR for a sample benchmark of two LLM-based defense agents. A Python controller connects five external components (Fig.~\ref{fig:architecture}): the GOAD lab~\cite{goad}, an eight-host Active Directory environment on Azure; NodeZero, Horizon3's commercial pentesting platform, as the autonomous attacker; Redshift, which stores NodeZero's attack telemetry (events, paths, weaknesses, credentials); Microsoft Defender XDR, whose detection and response telemetry we query through the Advanced Hunting API; and a proposer wrapper for the defense agent under test, which receives the telemetry and returns policies to deploy. We later swap the EDR/XDR component for CrowdStrike Falcon and SentinelOne Singularity, reusing the same environment, attacker, and loop (Appendix~\ref{app:vendor-results}). We motivate our choices of environment, attacker, and EDR/XDR in Sections~\ref{sec:goad}, \ref{sec:nodezero}, and~\ref{sec:edrxdr}.

\begin{figure}[!t]
\centering
\resizebox{\columnwidth}{!}{%
\begin{tikzpicture}[
    node distance=0.22cm and 0.6cm,
    host/.style={draw, rounded corners=2pt, minimum width=2.1cm,
                 minimum height=0.6cm, align=center, font=\scriptsize},
    jb/.style={host},
    domain/.style={draw, dashed, rounded corners=4pt, inner sep=3pt,
                   label={[font=\tiny\itshape, anchor=south, yshift=-1pt]north:#1}},
    forest/.style={draw, thick, rounded corners=6pt, inner xsep=5pt, inner ysep=10pt},
    trust/.style={{Stealth[length=2mm]}-{Stealth[length=2mm]}, densely dashed, thick, gray!70},
    trustlbl/.style={font=\tiny, fill=white, inner sep=1pt, text=gray!70},
    conn/.style={-{Stealth[length=2mm]}, thin, gray!50},
    connlbl/.style={font=\tiny, fill=white, inner sep=1pt, text=gray!50},
]


\node[host]  (dc01)  {dc01 {\tiny(DC)}\\[-1pt]{\tiny kingslanding}};
\node[host, below=of dc01] (srv01) {srv01 {\tiny(Server)}\\[-1pt]{\tiny the-eyrie}};
\node[host, below=of srv01] (ws01)  {ws01 {\tiny(WS)}\\[-1pt]{\tiny casterlyrock}};

\node[domain={sevenkingdoms.local}, fit=(dc01)(srv01)(ws01)] (sev) {};

\node[host, right=1.4cm of dc01]  (dc02)  {dc02 {\tiny(DC)}\\[-1pt]{\tiny winterfell}};
\node[host, below=of dc02] (srv02) {srv02 {\tiny(Server)}\\[-1pt]{\tiny castelblack}};

\path let \p1=(ws01.south) in node[minimum height=0pt, inner sep=0pt]
    at (dc02 |- \p1) (north_anchor) {};
\node[domain={north.sevenkingdoms.local}, fit=(dc02)(srv02)(north_anchor)] (north) {};

\node[forest, fit=(sev)(north)] (f1) {};
\node[font=\tiny\bfseries, anchor=south west] at (f1.south west)
    {\,Forest 1};


\node[host, right=1.4cm of dc02]  (dc03)  {dc03 {\tiny(DC)}\\[-1pt]{\tiny meereen}};
\node[host, below=of dc03] (srv03) {srv03 {\tiny(Server)}\\[-1pt]{\tiny braavos}};
\node[host, below=of srv03] (lx01)  {lx01 {\tiny(Linux)}\\[-1pt]{\tiny dragonstone}};

\node[domain={essos.local}, fit=(dc03)(srv03)(lx01)] (ess) {};
\node[forest, fit=(ess)] (f2) {};
\node[font=\tiny\bfseries, anchor=south west] at (f2.south west)
    {\,Forest 2};

\draw[trust] (sev.east |- dc01.south) -- (north.west |- dc02.south)
    node[trustlbl, midway, below=1pt] {parent--child};
\draw[trust] (f1.east) -- (f2.west)
    node[trustlbl, midway, above=1pt] {external trust};

\node[jb, below=0.45cm of $(f1.south)!0.5!(f2.south)$] (jb)
    {jumpbox};

\draw[conn] (jb) -- (f1.south);
\draw[conn] (jb) -- (f2.south);
\node[connlbl, right=0.1cm of jb] {WinRM / SSH};

\end{tikzpicture}}
\caption{GOAD lab network topology. Eight hosts span two Active Directory forests and three domains on a shared Azure subnet. Dashed arrows are trusts: parent--child within Forest~1, external between the two forest roots. The jumpbox is the WinRM/SSH relay and NodeZero runner.}
\label{fig:topology}
\end{figure}

\subsubsection{Environment: GOAD Lab}\label{sec:goad}

We use the GOAD lab as the test environment, extended to eight hosts: seven Windows and one Linux (Fig.~\ref{fig:topology}). The topology spans two Active Directory forests (\texttt{sevenkingdoms.local} and \texttt{essos.local}) connected by an external trust between root domains, with three domains in total (a child domain, \texttt{north.sevenkingdoms.local}, extends the first forest) and one domain controller each. A jumpbox VM on the same Azure virtual network serves as the WinRM relay and NodeZero runner.

This topology is the minimum sufficient configuration that exercises the attack surface the Microsoft Defender XDR policy catalog targets. The two-forest cross-trust forces NodeZero to perform cross-domain credential theft and trust traversal rather than reaching every host from any compromised one. The three member servers (one per domain) contribute application-service attack surface between the foothold and the domain controllers: Microsoft SQL Server (MSSQL), Internet Information Services (IIS), Exchange, Active Directory Certificate Services (ADCS), and Server Message Block (SMB) shares. This gives NodeZero service-level credential-theft and ADCS privilege-escalation targets. The Linux endpoint confirms that our framework correctly skips Windows-only policies on incompatible hosts. The single workstation (\texttt{ws01}) is the typical initial-access target and the only host where browser- and Office-oriented controls, such as some Attack Surface Reduction (ASR) rules and exploit-protection mitigations, apply. We treat the lab as a controlled testbed for the policy-selection problem, not as a full enterprise emulation. At eight VMs, the lab is reproducible at moderate cloud cost, a NodeZero pentest completes in roughly one to two hours, and our framework can instrument and observe every host individually.

\begin{figure*}[b]
\centering
\resizebox{\textwidth}{!}{%
\begin{tikzpicture}[
    node distance=0.4cm and 0.55cm,
    stage/.style={draw, rounded corners=2pt, minimum width=1.5cm, minimum height=0.65cm, align=center, font=\scriptsize},
    pf/.style={stage},
    su/.style={stage},
    lp/.style={stage},
    td/.style={stage},
    arr/.style={-{Stealth[length=2mm]}, semithick},
    lane/.style={font=\scriptsize\bfseries, anchor=east},
    loopback/.style={-{Stealth[length=2mm]}, semithick, rounded corners=4pt},
]

\node[lane] at (-0.3, 2.4) {Preflight};
\node[lane] at (-0.3, 1.2) {Setup};
\node[lane] at (-0.3, 0.0) {Loop};
\node[lane] at (-0.3,-1.2) {Teardown};

\node[pf] (p1) at (1.2, 2.4) {reachability};
\node[pf, right=of p1] (p2) {MDE status};
\node[pf, right=of p2] (p3) {credentials};
\node[pf, right=of p3] (p4) {runner};
\node[pf, right=of p4] (p5) {queue};
\draw[arr] (p1)--(p2); \draw[arr] (p2)--(p3); \draw[arr] (p3)--(p4); \draw[arr] (p4)--(p5);

\node[su] (s1) at (1.2, 1.2) {baseline};
\node[su, right=of s1] (s2) {provision};
\node[su, right=of s2] (s3) {export};
\node[su, right=of s3] (s4) {enrich};
\node[su, right=of s4] (s5) {build gym};
\draw[arr] (s1)--(s2); \draw[arr] (s2)--(s3); \draw[arr] (s3)--(s4); \draw[arr] (s4)--(s5);

\node[lp] (l1) at (1.2, 0.0) {propose};
\node[lp, right=of l1] (l2) {reconfig};
\node[lp, right=of l2] (l3) {wait};
\node[lp, right=of l3] (l4) {provision};
\node[lp, right=of l4] (l5) {export};
\node[lp, right=of l5] (l6) {enrich};
\node[lp, right=of l6] (l7) {reveal};
\node[lp, right=of l7] (l8) {compare};
\draw[arr] (l1)--(l2); \draw[arr] (l2)--(l3); \draw[arr] (l3)--(l4);
\draw[arr] (l4)--(l5); \draw[arr] (l5)--(l6); \draw[arr] (l6)--(l7); \draw[arr] (l7)--(l8);
\draw[loopback] (l8.south) -- ++(0,-0.4) -| (l1.south);

\node[td] (t1) at (1.2,-1.2) {reset};
\node[td, right=of t1] (t2) {MDE cleanup};
\draw[arr] (t1)--(t2);

\end{tikzpicture}}
\caption{Pipeline stage flow. Four stages: preflight validates infrastructure, setup runs the baseline pentest, the loop iterates policy proposals through deployment and measurement, and teardown resets the lab. The loop repeats until a stop criterion fires.}
\label{fig:pipeline}
\end{figure*}

\subsubsection{Attacker: NodeZero}\label{sec:nodezero}
We select NodeZero, Horizon3's commercial autonomous pentester, which chains multi-step attacks to mimic real attacker behavior~\cite{nodezero}. As discussed in Section~\ref{sec:background}, interpreting chained attacks has been cited as a limitation of commercial EDR. NodeZero is plugged into the framework via its H3 CLI connector.

\subsubsection{Commercial EDR/XDR}\label{sec:edrxdr}
We chose Microsoft Defender XDR because it is the market-share leader among commercial EDR products and its Advanced Hunting query surface preserves per-event data at a granularity sufficient for raw-telemetry correlation. To demonstrate the advantage of our open architecture design, we also run the framework on two other major commercial products, CrowdStrike Falcon and SentinelOne Singularity. Each exposes a different telemetry surface that we correlate with a vendor-specific method: CrowdStrike Falcon Insight detections~\cite{falcon-insight} and SentinelOne Storyline event graphs~\cite{sentinelone-sing}.

\subsection{Four Stages of Closed-Loop Evaluation}\label{sec:pipeline}
The experiment controller executes four stages (Fig.~\ref{fig:pipeline}) for experiment runs: Preflight, Setup, Loop, and Teardown.

\subsubsection{Preflight}\label{sec:preflight}
Running a closed-loop trial against live infrastructure is expensive. Each experiment run can take up to four hours and a failed experiment wastes that time and leaves cloud-side state changes that must be reset before retrying. The preflight stage (Table~\ref{tab:preflight}) catches the most common failures before the experiment starts, running seven automated checks in parallel across host reachability, MDE onboarding state, credentials, and infrastructure health. 

\begin{table}[h]
\renewcommand{\arraystretch}{1.1}
\centering
\caption{Preflight check phases.}
\label{tab:preflight}
\begin{tabular}{llc}
\toprule
Phase & Check & Auto-fix \\
\midrule
Host reachability & WinRM/SSH probe    & Yes \\
MDE onboarding    & Health, risk score & Partial \\
AWS SSO           & STS identity       & No \\
Azure CLI         & Account + RG       & No \\
Runner disk       & Free space         & Yes \\
Redshift DNS      & DNS resolution     & No \\
Runner queue      & Active pentests    & No \\
\bottomrule
\end{tabular}
\end{table}

Blockers (unreachable hosts, offboarded agents, expired credentials) fail the preflight; warnings (elevated risk score, stale alerts from prior runs) are surfaced but do not block. An auto-fix mode can remediate a subset of blockers without human intervention: restarting unreachable VMs via the Azure CLI, pruning the jumpbox Docker cache, and running the teardown cleanup to clear stale alerts. 

\subsubsection{Setup}\label{sec:setup}
Setup establishes a uniform baseline configuration across all hosts, runs a pentest to measure the starting state, and builds an initial attack graph from the resulting offensive and defensive telemetry that the proposer will reason over in subsequent rounds.

\subsubsection{Loop}\label{sec:loop}
The experiment controller executes the closed-loop policy-tuning protocol in the loop stage. Each iteration proceeds through eight stages until the stop criterion is met: 
\begin{enumerate}
    \item \textbf{Propose}: Proposer receives the candidate policy list and the enriched telemetry from the prior round, and returns a set of new policy selections.
    \item \textbf{Reconfig}: Policy selections are deployed to the appropriate host(s).
    \item \textbf{Wait}: Framework waits until the VM reboots complete and policy deployments have propagated.
    \item \textbf{Provision}: Autonomous attacker (NodeZero) begins a pentest.
    \item \textbf{Export}: Offensive telemetry is pulled from the data store and the defensive telemetry is extracted using the vendor's API configuration.
    \item \textbf{Enrich}: Offensive and defensive telemetry streams are tagged and joined into a per-event table.
    \item \textbf{Reveal}: Ground-truth attack graph is reconstructed from the fused offensive and defensive telemetry and made available alongside the enriched table for the next round and for post-experiment metrics.
    \item \textbf{Compare}: Per-iteration testing metrics are computed against the prior round.
\end{enumerate}

After every pentest, the controller pulls three data sources: NodeZero's attack telemetry from Redshift; Microsoft Defender XDR's telemetry from the Advanced Hunting API (replaceable with CrowdStrike Falcon and SentinelOne Singularity APIs); and on-host Windows event logs pulled per-host via WinRM (Windows Defender Application Control (WDAC) enforcement events, Firewall block events, and Windows Filtering Platform packet-drop events), which fire locally but are not reliably forwarded to Advanced Hunting. We correlate offensive and defensive telemetry using a deterministic pipeline:

\begin{enumerate}
    \item \textbf{Static tags}: Attaches MITRE technique and tactic IDs to each NodeZero event.
    \item \textbf{Bypass verdict}: Flags steps where a deployed policy should have intervened but the step succeeded.
    \item \textbf{Blast radius}: Propagates downstream impact (hosts compromised, credentials harvested) onto the parent step.
    \item \textbf{Vendor policies}: Annotates each step with the set of catalog policies whose MITRE techniques overlap.
    \item \textbf{Attack context}: Attaches the active credentials and implant sessions from the pentest export to each step.
    \item \textbf{Threat characterization}: Classifies each step into reconnaissance, credential theft, lateral movement, privilege escalation, or objective.
    \item \textbf{EDR correlation}: Matches each NodeZero step to the merged Microsoft Defender XDR timeline on the same host, using process-lineage structure as the primary join and time only as a coarse filter (detailed below).
\end{enumerate}

Because NodeZero and Defender clocks do not align and EDR reporting latency is large, we do not match on exact time. The primary join is structural. For each step we walk the process family tree on that host from NodeZero-attributed roots. We never trace through shared operating-system processes such as \texttt{lsass.exe} or \texttt{svchost.exe}, so a match rests on a process lineage, with time only as a coarse filter. We use a five-minute host-scoped window for the lowest-confidence tier, the one with only a time signal, and a $\pm$300-second window to bundle a matched event's neighbors. Both are far narrower than the eight-to-24-minute delay of throttled ASR events. Microsoft reports those events once per process per hour, so we do not reconstruct the suppressed occurrences and treat per-policy block counts as~lower~bounds.

The same structural approach carries to the other two vendors, each using its own lineage signal. For CrowdStrike, we attribute an event to NodeZero when it sits in one of its process trees. We expand the tree from attacker-rooted anchors by parent-process ID rather than CrowdStrike's native \texttt{TreeId} field, which is populated in under 1\% of events and too sparse to rely on. For SentinelOne, we seed from the runner's outbound connections and follow the Storyline graph. Inbound contact back to the runner tags only that single event; we never let it root a new storyline, which would attribute the host's unrelated activity to NodeZero. Both apply the same content check and one-to-one event assignment as the Defender path.

The experiment controller reconstructs the ground-truth attack graph from the correlated offensive and defensive data, then passes it to the proposer.

\subsubsection{Teardown}\label{sec:teardown}
After every experiment, the teardown stage resets each host to its baseline configuration and clears accumulated cloud-side state. The host reset removes all deployed policies (ASR rules, WDAC, firewall changes, etc.) so the lab returns to its pre-experiment state. The cloud cleanup resolves open alerts and active incidents on lab hosts, releases any device isolation, and resets host risk scores in the Microsoft Defender XDR portal, so the next pentest starts from a clean state. Without this cleanup, alerts from prior runs accumulate and escalate hosts to high risk, causing Microsoft Defender XDR's automatic attack disruption engine to auto-disrupt subsequent pentests. In early trials, automatic attack disruption contained lab user accounts, isolated machines, and killed NodeZero's credential-theft and lateral-movement actions before they could produce telemetry, resulting in near-empty pentest results across multiple consecutive runs. User containment release is the one exception: it is not yet exposed via API and requires manual action in the Microsoft Defender XDR portal. CrowdStrike and SentinelOne required only the host and policy reset; because neither disrupted runs autonomously, no equivalent cloud cleanup was needed.

\section{Sample Benchmark}
We run a sample benchmark comparing two LLM backbones for the autonomous defense agent on Microsoft Defender XDR using our evaluation framework. 

\subsection{Agent Architecture}
The two LLMs share an agentic architecture adapted from prior work on autonomous tuning of commercial EDR~\cite{prinos} with extensions to the defender action space and budget modeling.
\subsubsection{Policy catalog}
The defender selects from a finite catalog of Microsoft Defender XDR policies. Table~\ref{tab:catalog} summarizes the action space.

\begin{table}[h]
\renewcommand{\arraystretch}{1.1}
\centering
\caption{Microsoft Defender XDR policy catalog.}
\label{tab:catalog}
\footnotesize
\begin{tabular}{lcp{2.4cm}}
\toprule
Policy type & Ct. & Mechanism \\
\midrule
ASR rules       & 19 & Per host per GUID \\
Folder access   & 1  & MpPreference \\
PUA protection  & 1  & MpPreference \\
Firewall        & 1  & NetFirewallProfile \\
LSA protection  & 1  & Registry write \\
WDAC            & 1  & XML + CIM refresh \\
Audit policy    & 1  & auditpol.exe \\
Exploit prot.   & 1  & ProcessMitigation \\
\bottomrule
\end{tabular}
\end{table}

Three additional policy types are recognized but deferred, each for a different reason: credential guard (the lab VMs do not meet the hardware prerequisites~\cite{cred-guard-prereqs}), EDR block mode (no deployment API without Intune E5 licensing), and identity protection (tenant-wide, so it cannot be scoped per host). Controls that protect against attack vectors absent from the lab are also excluded: SmartScreen and web content filtering (no interactive users browsing the web), device control and hardware installation restrictions (no removable media on cloud-hosted VMs), and BitLocker (data-at-rest encryption with no bearing on network-level attack paths). Cloud-delivered protection and cloud sample submission are disabled because Microsoft's cloud-side detection intelligence changes independently of our policy deployments, introducing an uncontrolled variable.

The agent's action space is the entire catalog in Table~\ref{tab:catalog}, available every round. The agent proposes only policies whose techniques overlap the weaknesses NodeZero actually exercised. The narrow set it converges on in our study (two ASR rules and LSA protection) reflects the credential-theft and lateral-movement attack surface of the GOAD lab, not a cap on the action space. We expect a different attack surface to exercise a different subset. Separately, the catalog policies deferred earlier for licensing and hardware reasons narrow what is available, but that too is a lab constraint, not a limit of the framework.

\subsubsection{Policy modes}
At any given iteration, each Microsoft Defender XDR policy can be set to one of three modes: \emph{block}, which enforces the control; \emph{log-only}, which audits without enforcement to expand visibility at low operational risk; and \emph{disable}, the default state before any deployment. Not all policies support all three modes. ASR rules, firewall, controlled folder access, WDAC, and Potentially Unwanted Application (PUA) protection use the full set. Audit policy supports log-only and disable but has no block concept, since subcategories are either audited or not. ASR rules additionally define a vendor ``Warn'' mode that blocks with a user-bypass prompt~\cite{ms-asr}; we collapse Warn into block because the lab has no interactive users. Mode transitions carry asymmetric costs, shown in Table~\ref{tab:transitions}, calibrated to Microsoft's documented deployment cadence~\cite{microsoft_asr_faq,microsoft_recs}.

\begin{table}[h]
\renewcommand{\arraystretch}{1.1}
\centering
\caption{Mode transition costs.}
\label{tab:transitions}
\footnotesize
\begin{tabular}{lrl}
\toprule
Transition & Cost & Rationale \\
\midrule
log-only $\to$ block     & 0.20 & 30-day audit + FP triage \\
off $\to$ block     & 0.33 & Cold-start enforcement \\
block $\to$ log-only     & 0.04 & Rollback to audit \\
block $\to$ off     & 0.02 & Emergency disable \\
off $\to$ log-only       & 0.02 & Trivial \\
log-only $\to$ off       & 0.02 & Trivial \\
\bottomrule
\end{tabular}
\end{table}

\begin{table*}[t]
\centering
\caption{Per-round policy selections. LSASS = ASR credential-theft rule (9e6c4e1f); PSExec = ASR PSExec/WMI rule (d1e49aac); LSA = LSA protection. Cost is the transition-cost budget spent that round ($B = 0.66$ per round).}
\label{tab:per-round-policies}
\scriptsize
\setlength{\tabcolsep}{4pt}
\begin{tabular}{cc|p{4.8cm}r|p{4.8cm}r}
\toprule
 & & \multicolumn{2}{c|}{Claude Sonnet 4.6}
   & \multicolumn{2}{c}{Cisco Foundation-Sec-8B} \\
\cmidrule(lr){3-4} \cmidrule(lr){5-6}
Pair & Iter & New Picks & Cost & New Picks & Cost \\
\midrule
1 & 1 & LSASS:block, LSA:block & 0.66 & LSA:disable, PSExec:block & 0.33 \\
 & 2 & PSExec:block & 0.33 & LSASS:block, LSA:block & 0.66 \\
 & 3 & LSA:block, LSASS:log-only, PSExec:log-only & 0.41 &  &  \\
 & \textit{Final} & \textit{LSA:block, LSASS:log-only, PSExec:log-only} & & \textit{LSA:block, LSASS:block, PSExec:block} & \\
 & \textit{Stop} & \textit{no candidates} & & \textit{no candidates} & \\
\midrule
2 & 1 & LSASS:block, LSA:block & 0.66 & LSA:disable, PSExec:log-only & 0.02 \\
 & 2 & PSExec:block & 0.33 & PSExec:block, LSASS:log-only & 0.22 \\
 & 3 &  &  & LSASS:block, LSA:block & 0.53 \\
 & \textit{Final} & \textit{LSA:block, LSASS:block, PSExec:block} & & \textit{LSA:block, LSASS:block, PSExec:block} & \\
 & \textit{Stop} & \textit{no candidates} & & \textit{no candidates} & \\
\midrule
3 & 1 & LSASS:block, LSA:block & 0.66 & LSA:disable, PSExec:block & 0.33 \\
 & 2 &  &  & LSASS:block, LSA:block & 0.66 \\
 & \textit{Final} & \textit{LSA:block, LSASS:block} & & \textit{LSA:block, LSASS:block, PSExec:block} & \\
 & \textit{Stop} & \textit{no candidates} & & \textit{no candidates} & \\
\midrule
4 & 1 & LSASS:block, LSA:block & 0.66 & LSA:disable & 0.00 \\
 & 2 & PSExec:block & 0.33 &  &  \\
 & \textit{Final} & \textit{LSA:block, LSASS:block, PSExec:block} & & \textit{LSA:disable} & \\
 & \textit{Stop} & \textit{no candidates} & & \textit{no candidates} & \\
\bottomrule
\end{tabular}
\end{table*}

The per-round budget $B = 0.66$ limits how many transitions fit in one iteration: three cheap upgrades fit ($0.20 \times 3$), two cold-starts fit exactly ($0.33 \times 2$), but mixing a cold-start into a three-upgrade round exceeds the budget. The asymmetry (upgrades cost engineer-weeks, downgrades cost engineer-days) makes rollback a tractable action.

\textbf{Game Value and Budget Gating.} Before deployment, a deterministic gate checks that (a) the transition cost fits within $B$, (b) the policy set does not increase the attacker's expected gain (game value), and (c) it does not contain off-catalog (hallucinated) policies. If all three hold, the picks deploy as-is. If any fails, the next-best cost-feasible, non-worsening option is deployed. If none exists, nothing is deployed and the experiment stops, as detailed in Section~\ref{sec:stop}.

\subsubsection{LLM backbones}
For a frontier general-purpose model we use Anthropic Claude Sonnet 4.6 which has performed well on cybersecurity benchmarks~\cite{sonnet_model_card}. We compare against Cisco Foundation-Sec-8b~\cite{cisco_model_card}, a specialized, open-weight security model developed by a team of researchers at Cisco. Appendix~\ref{app:repro} includes implementation details.

\subsection{Benchmark Setup}
We run four paired experiments. Each pair shares one baseline pentest; both proposers then run their closed-loop trials from that common starting point. Table~\ref{tab:per-round-policies} summarizes all trials. This two-agent comparison runs on Microsoft Defender XDR only; we use CrowdStrike Falcon and SentinelOne Singularity to test whether the lessons hold across vendors, not to compare agents. Across all pairs, Claude Sonnet 4.6 and Cisco Foundation-Sec-8B each completed eight deployment iterations, with every trial terminating when the candidate filter returned no remaining policies. In six of the eight trials, both proposers converged to the same three-policy set (the LSASS credential-theft ASR rule, the PSExec/WMI ASR rule, and LSA protection), typically within one to three deployment rounds.

Table~\ref{tab:per-round-policies} details the round-by-round policy selections per proposer. Claude Sonnet 4.6 consistently led with the LSASS and LSA rules at full budget (0.66) in its first round, adding the PSExec/WMI rule in a subsequent round. Cisco Foundation-Sec-8B showed more variable behavior: in Pairs~1 and~3 it selected PSExec first (skipping the LSASS rule entirely in the first round), in Pair~2 it began with a low-cost log-only deployment before upgrading, and in Pair~4 it proposed only LSA:disable.

These trajectories also yield different final configurations in three of the four pairs (Table~\ref{tab:per-round-policies}), and even where the two agents converge to a similar rule set, their deployment paths differ. The framework records these path and configuration differences. An aggregate benchmark that reports only the final configuration would miss them.

\subsubsection{Hyperparameters}\label{sec:hyperparameters}
All parameters are configurable per experiment via a YAML file. The convergence model and stop-criterion parameters are adapted from~\cite{prinos}. For the trials, both proposers use identical values detailed in Table~\ref{tab:hyperparams}.

\begin{table}[h]
\renewcommand{\arraystretch}{1.1}
\centering
\caption{Hyperparameters.}
\label{tab:hyperparams}
\begin{tabular}{lcl}
\toprule
Parameter & Value & Description \\
\midrule
$B$            & 0.66      & Per-round cost budget \\
$R$                        & 0.05      & Kalman measurement noise \\
$\lambda$                  & 1.0       & Lyapunov observer weight \\
$\varepsilon_\text{innov}$ & 0.05      & Innovation threshold \\
$\varepsilon_V$            & $10^{-4}$ & $\Delta V$ threshold \\
max\_iterations            & 10        & Hard cap on rounds \\
\bottomrule
\end{tabular}
\end{table}

The per-round cost budget $B = 0.66$ is derived from the transition cost model (Table~\ref{tab:transitions}). The Kalman measurement noise $R = 0.05$ reflects high trust in each pentest observation, since pentest outcomes are observed directly, not simulated; higher $R$ would slow belief convergence and require more iterations. The Lyapunov weight $\lambda = 1.0$ gives equal importance to the game value $S$ and the defender confidence $\theta$. The ten-iteration hard cap is a practical constraint: each iteration includes a full pentest (one to two hours), so a ten-round experiment runs for roughly 10--20~hours.

\subsubsection{Stop criteria}\label{sec:stop}

The loop terminates on the first of these triggers:

\begin{enumerate}
    \item \textbf{Empty proposal}: the proposer recommends zero policies. One more round confirms before stopping.
    \item \textbf{Saturated}: the candidate filter returns zero candidates; every compatible policy is already deployed.
    \item \textbf{Max iterations}: the ten-round hard cap.
    \item \textbf{Stage failure}: any stage raises an error (fail-fast).
\end{enumerate}

\subsection{Telemetry Analysis}
To validate the evaluation framework rather than the defensive agents, we analyze the raw offensive and defensive telemetry manually after each run, assisted by Claude Opus 4.7 (1M context). Errors in the deterministic tagging and correlation pipeline (Section~\ref{sec:loop}) could affect the agents' recommendations during a run, but our manual analysis keeps them out of the reported findings. For reproducibility, deterministic Python scripts compute aggregate figures.

\section{Results and Lessons Learned}
\label{sec:results}
We ran four paired closed-loop experiments, each pairing a Claude Sonnet 4.6 trial with a Cisco Foundation-Sec-8B trial on a shared baseline. Across 20 pentests and 16 deployment iterations, we analyzed about 2,388 raw NodeZero offensive events and 207k Microsoft Defender XDR defensive alerts, deduplicated across shared baselines.

To evaluate generality beyond a single vendor, we also ran the framework on CrowdStrike Falcon and SentinelOne Singularity, changing only the EDR and its correlation method: the GOAD lab, NodeZero attack template, and defense-agent loop were identical. On each vendor we ran a single closed-loop trial with the Claude Sonnet 4.6 proposer, which ran to three pentests on CrowdStrike Falcon and five on SentinelOne Singularity. Their action spaces are cataloged in Appendix~\ref{app:action-spaces} and per-run results are in Appendix~\ref{app:vendor-results}.

These experiments surface the EDR-evaluation challenges from Section~\ref{sec:background} and highlight the limitations of existing approaches that rely on abstract defensive measures, simulation, emulation, or open-source EDRs. Each of the three lessons has its own subsection below. Within each, we separate the finding specific to Microsoft Defender XDR, the broader cross-EDR behavior, and the methodological lessons learned. We then report attack-centric outcomes and close with limitations and open questions.

\subsection{Commercial EDR Telemetry Is Not Designed for Measuring Policy Effectiveness} 
\subsubsection{Microsoft-Defender-XDR-specific finding}
ASR events in the Advanced Hunting \texttt{DeviceEvents} table are throttled to unique processes seen every hour, and stamped with the first occurrence within the hour, as documented by Microsoft~\cite{asr_monitor}. Attack disruption's \texttt{ContainedUser*} events are not throttled and retain second-level precision. Microsoft also documents that Microsoft Defender XDR's incident-graph and alert-evidence surfaces (\texttt{AlertInfo}, \texttt{AlertEvidence}) carry different per-event information than \texttt{DeviceEvents}. Though well-documented by Microsoft, these design choices and their effect on defense-configuration evaluation are absent from the literature.

The number of times a rule fired is not the number of times it caught the attacker. The two ASR rules break that assumption in opposite directions: the LSASS rule's count is inflated by benign system processes, and the PSExec/WMI rule's is deflated by throttling. The agents deployed two ASR rules in every experiment: the LSASS credential-theft rule (GUID: 9e6c4e1f-7d60-472f-ba1a-a39ef669e4b2), targeting credential access (T1003.001), and the PSExec/WMI execution rule (GUID: d1e49aac-8f56-4280-b9ba-993a6d77406c), targeting command execution (T1047 / T1569.002). Across all eight experiments, the LSASS credential-theft rule fired on 55 of 370 NodeZero credential-dump actions (15\%) and the PSExec/WMI execution rule fired on 47 of 66 NodeZero WMI/exec actions (71\%). Neither rule logged a block within $\pm$60 seconds of a matching NodeZero action; ASR logged its block a median of eight to 24 minutes after the action. The ASR vendor documentation accounts for this latency, but not whether either rule actually caught the attacker. Closer inspection showed the 55 LSASS-rule events fired predominantly on benign Windows system activity, not attacker tooling: 38 (69\%) initiated by \texttt{svchost.exe}, four (7\%) by Azure platform agents (\texttt{WindowsAzureGuestAgent.exe}, \texttt{WaAppAgent.exe}), and the remainder by system processes (\texttt{cmd.exe}, \texttt{powershell.exe}, \texttt{SenseIR.exe}, \texttt{lsass.exe}). The 47 PSExec/WMI events fired primarily on \texttt{WmiPrvSE.exe} PIDs (19 events, 40\%) that NodeZero spawned per WMI command, with the remainder on \texttt{cmd.exe} (12 events) and NodeZero implant binaries. Because ASR throttles to a single reported event per hour per PID, this can undercount how often the PSExec/WMI rule actually fired on NodeZero processes, so the true rate may be higher than advanced hunting shows.

To assess the magnitude of underreporting on Microsoft Defender XDR, we cross-referenced its named-block telemetry with NodeZero's per-action outcomes (Success, Failure, or Unknown, with a \texttt{failure\_type} discriminator on failures). Across 2,388 NodeZero actions, 15 (0.6\%) were failed actions whose payloads were retrieved but whose execution was prevented. Of these 15 confirmed interventions, 13 (87\%) have a matching Microsoft Defender XDR named-block within $\pm$300 seconds on the same host. NodeZero recorded the other two as blocked from the attacker side, but Microsoft Defender XDR's \texttt{DeviceEvents} stream never surfaced them as \texttt{ContainedUser*} or \texttt{Asr*Blocked} action types. So 13\% of confirmed interventions leave no named-block trace in the defensive telemetry.

\subsubsection{Cross-EDR generalization} We extended the framework to CrowdStrike Falcon and SentinelOne Singularity to test whether these patterns held on other vendors. Unlike Microsoft Defender XDR, CrowdStrike Falcon enforced nothing on its own: its baseline run, with no policy deployed, produced zero blocks, whereas Microsoft Defender XDR's automatic attack disruption layer contained users autonomously even at baseline. Once the agent deployed a credential-dumping policy, CrowdStrike Falcon's enforcement covered just part of the technique: it stopped 6 LSASS-access attempts but logged 32 registry-based credential accesses without blocking them (Appendix~\ref{app:vendor-results}). We inspected detections individually to aggregate enforcements since CrowdStrike Falcon records block status only within each detection. 

Unlike Microsoft Defender XDR and CrowdStrike Falcon, which record enforcement per action, SentinelOne Singularity records per threat, and one mitigation can span several NodeZero actions. Across its two protect-mode rounds, it mitigated 24 threats in the first and 11 in the second(Appendix~\ref{app:vendor-results}). This threat-level grouping complicated correlation, so we could not tie any of the 35 mitigations to a specific NodeZero action. We find that each product records defensive outcomes for its own SOC workflow (CrowdStrike Falcon Insight investigations~\cite{falcon-insight}, SentinelOne Storyline event graphs~\cite{sentinelone-sing}, Microsoft Defender XDR \texttt{AlertInfo}/\texttt{AlertEvidence})~\cite{defender-alert}, not for measuring policy effectiveness.

\begin{table*}[!t]
\centering
\caption{Block attribution across all post-baseline pentests. ASR: operator-deployed ASR rules. Contain: all \texttt{ContainedUser*} action types (SMB, RPC, and RDP-session containment) from automatic attack disruption.}
\label{tab:asr-vs-air}
\begin{tabular}{cl rrr rr}
\toprule
Pair & Proposer & Pentests & $\sum$ASR & $\sum$Contain
     & ASR\% & Contain\% \\
\midrule
1 & Claude Sonnet 4.6 & 3 & 14 & 794 & 1.7 & 98.3 \\
 & Cisco Foundation-Sec-8B & 2 & 12 & 607 & 1.9 & 98.1 \\
\cmidrule{2-7}
2 & Claude Sonnet 4.6 & 2 & 5 & 762 & 0.7 & 99.3 \\
 & Cisco Foundation-Sec-8B & 3 & 3 & 1563 & 0.2 & 99.8 \\
\cmidrule{2-7}
3 & Claude Sonnet 4.6 & 1 & 5 & 2 & 71.4 & 28.6 \\
 & Cisco Foundation-Sec-8B & 2 & 6 & 17 & 26.1 & 73.9 \\
\cmidrule{2-7}
4 & Claude Sonnet 4.6 & 2 & 6 & 36 & 14.3 & 85.7 \\
 & Cisco Foundation-Sec-8B & 1 & 0 & 926 & 0.0 & 100.0 \\
\bottomrule
\end{tabular}
\end{table*}

\subsubsection{Lessons learned and recommendations} Our findings corroborate reports that EDR fails to block or log most attacks~\cite{karantzas}. Our per-policy view adds a caveat: some of that apparent failure is a measurement artifact, not a true enforcement gap. The blocks visible in standard telemetry do not match the number the EDR fired, and that gap varies by rule. Vendor defaults such as ASR throttling can make a defensive configuration look weaker than it is. An aggregate metric like ``0\% of actions blocked within 60 seconds'' is ambiguous: it can mean a rule fired on the wrong processes (a rule-effectiveness gap), or fired on the right processes but logged its block outside the 60-second window (a telemetry-granularity gap). Differentiating these requires per-initiating-process attribution and awareness of per-PID throttling. The mechanism differs by vendor: Microsoft Defender XDR throttles repeated events, CrowdStrike Falcon records blocks only per detection, and SentinelOne Singularity groups outcomes by threat. None expose per-policy effectiveness directly. A finer-grained view recovers it on Microsoft Defender XDR and CrowdStrike Falcon, but SentinelOne Singularity records nothing below the threat level, preventing recovery. We recommend decomposing per-policy reports rather than trusting aggregates, scoping each claim to the telemetry surface it was measured on, and cross-checking against an independent attacker-side signal.

Silent-bypass rate is not fixed: it moves with how we define an EDR-evaded action, and cannot be measured the same way across vendors. An EDR-evaded action has two definitions: block-only evasion aggregates actions that pass without an enforced block, and detection evasion filters further to actions that also receive no detection (raw telemetry never counts under either). Across the EDR-observable actions (excluding network-layer actions without sensors), all three vendors leave most unblocked: CrowdStrike Falcon 99--100\%, SentinelOne Singularity 91.4-100\%, and Microsoft Defender XDR 94.6--100\%. Filtering further to undetected actions changes only SentinelOne Singularity, dropping it to 68--81\%; CrowdStrike Falcon and Microsoft Defender XDR barely change (Appendix~\ref{app:silent-bypass}). Measurability also differs by product. On CrowdStrike Falcon and SentinelOne Singularity, we separate a named detection from raw telemetry. On Microsoft Defender XDR, where the advanced-hunting timeline does not separate a named detection from telemetry, we pair it with the \texttt{AlertInfo} and \texttt{AlertEvidence} tables. These purge after about a month, so we could only export them for the automatic attack disruption runs (Section~\ref{subsec:air-variation}) and report the detection-evasion rate on those alone.

\subsection{Per-Policy Attribution Isolates the Defensive Agent's Effect}
\subsubsection{Microsoft-Defender-XDR-specific finding} Commercial EDRs embed autonomous components that act alongside operator-configured policies. To measure a defense agent's effect, we must separate the blocks its imposed policies caused from those the vendor's autonomous layer produced on its own. We show how our framework does this for Microsoft Defender XDR's automatic attack disruption. The underreporting artifacts from the previous section likely still apply to the block rates here.

ASR rules target specific process and API behaviors~\cite{ms-asr}: the agent's two ASR rules (Table~\ref{tab:per-round-policies}) block LSASS credential-memory access and PSExec/WMI process creation, covering credential access and command execution. Automatic attack disruption's \texttt{ContainedUser*} containment instead acts on session-level, multi-step activity; in our runs it fired on lateral-movement, collection, and DCSync techniques. The two layers cover disjoint MITRE tactics, and the agent can configure only the ASR layer. Per-policy decomposition is what separates the coverage produced by the EDR's autonomous decisions from the coverage produced by the policies the agent deployed.

Table~\ref{tab:asr-vs-air} quantifies this decomposition across all 16 post-baseline pentests. Operator-deployed ASR rules account for 1.8\% (Claude Sonnet 4.6) and 0.7\% (Cisco Foundation-Sec-8B) of all observed blocks; the vendor-autonomous automatic attack disruption layer accounts for 28.6--100\%. The wide range reflects variance in automatic attack disruption activity across pairs: in Pair~3 Claude Sonnet 4.6, automatic attack disruption fired only twice and ASR blocks constituted 71.4\% of the seven total blocks, an artifact of low activity in that pentest window, not higher ASR effectiveness. In contrast, Pairs~1 and~2 saw hundreds to thousands of automatic attack disruption events per trial, making operator-deployed ASR contributions negligible in proportion. We did not reset automatic attack disruption state between these pentest runs, so the automatic attack disruption counts reflect both per-run activity and containment carried over from earlier runs, not agent behavior; Section~\ref{subsec:air-variation} isolates that carryover from genuine per-run variation.

Per-policy attribution also surfaced false-positive behavior: the LSASS credential-theft ASR rule fired entirely on benign system processes rather than attacker tooling, as detailed in the preceding section.

\subsubsection{Cross-EDR generalization} In the CrowdStrike Falcon and SentinelOne Singularity runs, neither vendor blocked autonomously, so every enforced block is attributable to a policy the agent deployed. On CrowdStrike Falcon, deploying the Credential Dumping policy raised correlated telemetry from 16.6\% to 24.2\% and enforced 6 blocks: 2 process kills and 1 handle downgrade in the first round, then 1 and 2 in the convergence re-test, all tagged OS Credential Dumping (Appendix~\ref{app:vendor-results}). On SentinelOne Singularity, blocking appeared only after MitigationMode escalated from detect to protect, mitigating 24 threats and then 11, and the correlated event count fell from 174 under detect to 53 under protect(Appendix~\ref{app:vendor-results}). Microsoft Defender XDR's automatic attack disruption layer, by contrast, blocked independently of any operator policy, complicating policy attribution.

\subsubsection{Lessons learned and recommendations} The autonomous components of EDR leave us with several recommendations for similar studies. First, measure per-policy attribution at raw-telemetry granularity rather than aggregating at the EDR-platform level, to separate defense-agent blocks from autonomous-EDR blocks. The attack-disruption/ASR split on Microsoft Defender XDR, from 29\%/71\% to 100\%/0\% across pairs, was invisible in any aggregate metric. Second, under the default posture, report the agent's and the autonomous layer's contributions side by side as concurrent components. Third, to measure policy effectiveness alone, disable both autonomous layers: set the device group's AIR automation level to \texttt{No automated response} and add the experiment's assets to automatic attack disruption's exclusions. This isolates the agent's policies but no longer reflects the default posture operators run, so it complements the side-by-side reporting rather than replacing it. Finally, investigate the tradeoff between two action-space designs: one that includes tenant-level autonomous-layer controls, and one limited to policies not tied to any autonomous component.

\begin{table*}[t]
\centering
\caption{Baseline versus final iteration. Events: enriched/correlated event count. Det\%: share of events with any detection. Contain: \texttt{ContainedUserSmbFileOpenBlocked} + \texttt{ContainedUserRpcAccessBlocked} (all types in Table~\ref{tab:asr-vs-air}). $\Delta$: final $-$ baseline. Attack disruption confounds these measures, so they do not show security improvement (Appendix~\ref{app:attack-outcomes}).}
\label{tab:baseline-vs-final}
\scriptsize
\begin{tabular}{cl rrr rrr rr}
\toprule
 & & \multicolumn{3}{c}{Baseline}
   & \multicolumn{3}{c}{Final Iter}
   & & \\
\cmidrule(lr){3-5} \cmidrule(lr){6-8}
Pair & Proposer & Events & Contain & Det\%
     & Events & Contain & Det\%
     & $\Delta$Ev & $\Delta$Contain \\
\midrule
1 & Claude Sonnet 4.6 & 1790 & 991 & 98.9 & 1033 & 305 & 100.0 & -757 & -686 \\
 & Cisco Foundation-Sec-8B & 1790 & 991 & 98.9 & 560 & 303 & 86.1 & -1230 & -688 \\
\cmidrule{2-10}
2 & Claude Sonnet 4.6 & 1475 & 111 & 100.0 & 1560 & 387 & 89.7 & +85 & +276 \\
 & Cisco Foundation-Sec-8B & 1475 & 111 & 100.0 & 2282 & 86 & 100.0 & +807 & -25 \\
\cmidrule{2-10}
3 & Claude Sonnet 4.6 & 3479 & 2 & 96.0 & 4260 & 2 & 97.7 & +781 & +0 \\
 & Cisco Foundation-Sec-8B & 3479 & 2 & 96.0 & 3502 & 12 & 95.9 & +23 & +10 \\
\cmidrule{2-10}
4 & Claude Sonnet 4.6 & 3136 & 71 & 95.7 & 1780 & 34 & 97.8 & -1356 & -37 \\
 & Cisco Foundation-Sec-8B & 3136 & 71 & 95.7 & 3377 & 924 & 99.2 & +241 & +853 \\
\bottomrule
\end{tabular}
\end{table*}

\subsection{The EDR's Autonomous Behavior Varies Across the Evaluation Window}
\label{subsec:air-variation}
\subsubsection{Microsoft-Defender-XDR-specific finding} The EDR's autonomous behavior is not a controlled variable, which complicates our agent benchmarking. Policy-effectiveness measurements are conditional on environmental factors including users present, activity accumulation, account scoping of vendor-autonomous components, and threshold-tuning controls that operators set outside the experiment. The clearest example of these factors was carried-over containment state, which changed across a nine-day window with no configuration by the team or agent. On the same lab with the same no-policy configuration, \texttt{ContainedUserBlocked} output rose 8.9$\times$ between the 15~May baseline (111 events) and the 20~May baseline (991). This was not more enforcement per run: neither baseline cleared state, so containment from earlier automatic attack disruption activity carried into the later run. Rerunning the no-agent baseline with and without state reset confirms this (Appendix~\ref{app:air-carryover}). With reset (releasing every contained user and isolated device between runs), each run starts from zero containments. Without reset, contained users accumulate 0, 6, 12 and one host stays isolated into the third run, because contained users and isolated devices stay that way until manually cleared. New containments per run stay at 3 to 7 either way. Another effect of the carried-over containments is that one event type appears only in the later run: \texttt{ContainedUserRemoteDesktopSession*} fires when a contained user's Remote Desktop session is disconnected or stopped. Across all the runs with an agent-deployed policy, attack disruption contained primarily one account, \texttt{robb.stark}, the initial-foothold user NodeZero pivots into, and left every other account used by NodeZero alone (\texttt{administrator}, \texttt{goadmin}, \texttt{brandon.stark}, \texttt{arya.stark}, multiple domain users, machine accounts, and anonymous SMB sessions). In both respects, containment tracks the autonomous layer's own state and targeting rather than the agent's actions, requiring per-policy attribution to separate the two.

Table~\ref{tab:baseline-vs-final} compares each of the eight closed-loop trials' baseline to its final iteration and shows that automatic attack disruption counts vary substantially. Six of the trials converged to the same three-policy set (the LSASS and PSExec ASR rules plus LSA protection), yet the change in automatic attack disruption events ($\Delta$Contain) still varies widely, from $-688$ (Cisco Foundation-Sec-8B, Pair~1) to $+276$ (Claude Sonnet 4.6, Pair~2). The variation does not track the deployed policy. It comes from a mix of carried-over containment state and run-to-run variation in the autonomous layer. The metric amplifies this: each \texttt{ContainedUser*Blocked} event marks one blocked SMB or RPC action by a contained account, so the total tracks how much these accounts kept probing in a run. Our runs cannot determine how much of this variation comes from the defense agent, the autonomous attack disruption layer, or NodeZero's own run-to-run behavior, so we report it as a joint effect rather than a measured split. A mid-experiment change compounds it: starting with Pair~3, we manually cleared attack disruption state between runs, so Pairs~1 and~2 ran with carryover present and Pairs~3 and~4 without, which explains their lower baselines. The two trials with differing converged policy sets are Claude Sonnet 4.6 in Pair~3 ($\Delta$Contain~$=0$, LSASS and LSA only) and Cisco Foundation-Sec-8B in Pair~4 ($\Delta$Contain~$=+853$, LSA-disable only). Pair~4 ran with carryover cleared, so its +853 containment swing cannot come from prior state; under a single policy, it reflects the autonomous layer rather than the deployed policy. Therefore, cross-trial automatic attack disruption counts do not reliably measure the agent's effect.

\subsubsection{Cross-EDR generalization} CrowdStrike Falcon and SentinelOne Singularity run autonomous layers and hold persistent host-isolation state (SentinelOne Network Quarantine, CrowdStrike Network Containment). Neither fired in our runs, so we did not test whether their states accumulate. We leave that to future work, along with characterizing the autonomous layers, CrowdStrike Falcon's graded ML sliders, and multi-policy decomposition (both vendors deployed only one policy). Appendix~\ref{app:action-spaces} catalogs the full CrowdStrike Falcon and SentinelOne Singularity action spaces.

\subsubsection{Lessons learned and recommendations} As commercial EDRs deploy more autonomous AI/ML components alongside operator-configured policies, studying how these interact with defense agents becomes increasingly important. Fair evaluation across Microsoft Defender XDR runs is difficult because the vendor-autonomous layer was not a controlled variable. We had to clear accumulated alerts and release device isolation between Microsoft Defender XDR experiments to recover from cascaded auto-disruption. Reset removes carried-over state as a driver of between-run differences, but automatic attack disruption's runtime variability remains: even the three no-agent baseline runs under the reset condition (containment state cleared between runs) varied from run to run (automatic attack disruption blocks of 121, 60, and 106; Appendix~\ref{app:air-carryover}). We therefore do not attribute between-run differences to the agent alone; without reset, leftover state skews later runs further. Two limits bound this measurement: releasing a contained user has no vendor API, so it is manual; and three runs per condition prevent tight confidence intervals. We kept Microsoft Defender XDR at its default \texttt{Full} automation to stay representative of real enterprise posture; excluding the experiment's users, devices, and IP addresses from automatic attack disruption is possible and could isolate the agent's contribution more cleanly, which we recommend for future ablations.

Baseline characterization is difficult in an environment where autonomous components evolve. A single baseline pentest captures a snapshot that may not be representative several pentests later. We recommend that experiments and ablation studies vary not only the agent under test (LLM, prompt, optimizer), but also the dynamic-environment knobs vendors expose (AIR automation level, automatic attack disruption's exclusion lists and per-account scoping). In our experiments, we used the same NodeZero-injected credentials throughout, but the per-account-selectivity finding suggests our results may have differed had NodeZero pivoted through a different initial-foothold user. We recommend varying the account and context deliberately across runs and measuring the effect, rather than fixing them. If the decision algorithm uses expected policy effectiveness, we recommend updating the scores to reflect the enforcement observed across runs.

\subsection{Attack-Centric Outcomes}\label{subsec:attack-outcomes} Reducing defensive telemetry is not the same as reducing attack success; neither agent reduced what NodeZero achieved against each configuration. On Microsoft Defender XDR, the extent of compromise varied between runs: 4--7 hosts, 2--3 domain controllers and domains, 83--248 credentials, and 0--14 domain-admin credentials, while ADCS privilege escalation succeeded in every run (Appendix~\ref{app:attack-outcomes}). This variation did not track the deployed policy, so the agent does not appear to be its main driver. It reflects the autonomous attack disruption layer (Section~\ref{subsec:air-variation}) and NodeZero's own run-to-run differences, whose separate contributions our runs cannot cleanly quantify (Section~\ref{subsec:air-variation}). On CrowdStrike Falcon and SentinelOne Singularity, compromise was complete in every run regardless of policy: all 7 Windows hosts (the Linux endpoint was never compromised), all 3 domain controllers, all 3 domains, domain-admin credentials, and ADCS privilege escalation each time (Appendix~\ref{app:attack-outcomes}). Only the raw credential count varied, 109--171 on CrowdStrike Falcon and 71--118 on SentinelOne Singularity; the breadth of compromise did not change. The agents changed the defensive telemetry but not the extent of compromise, as discussed further in Appendix~\ref{app:attack-outcomes}.

\subsection{Limitations, Open Questions, and Future Work}\label{sec:scope-open}
Our framework and sample benchmarking study surface challenges specific to evaluating autonomous defense with commercial EDR tools, along with lessons and recommendations. Here we detail the limitations of our work and the open questions we leave to the practitioner and research communities.

\subsubsection{Limitations} Our results come from one lab (GOAD), one pentester (NodeZero), and one network topology, with Microsoft Defender XDR as the primary EDR and CrowdStrike Falcon and SentinelOne Singularity as a generalization probe. We designed the framework to extend to further EDR/XDR vendors, attackers, environments, and agent architectures. Our agents also deployed only a small subset of available policies. The transition-cost model and per-round budget ($B=0.66$) constrain the agents' available actions each round; we did not vary them, so some convergence behavior may be specific to these settings, and a sensitivity analysis remains future work. Runs are not identical: module ordering, credential reuse, and timing vary between them, and Microsoft Defender XDR's automatic attack disruption adds run-to-run variation (Table~\ref{tab:baseline-vs-final}). Commercial tooling also limits reproducibility; Appendix~\ref{app:repro} lists each component and its licensing or access requirement.

\subsubsection{Open questions} 
Several questions remain that this paper does not resolve. How far do our findings transfer? Open directions include other EDR vendors, lab topologies, and pentesters beyond the ones we study; the untested aspects of all three EDRs; and autonomous LLM-based attackers whose attack-path selection differs from NodeZero's. Commercial EDR vendors update their products continuously, so how should evaluation methodologies stay valid as vendor cloud-side classifiers are retrained and autonomous mechanisms are introduced or deprecated? More broadly, has evaluation focused too much on the attacker and the environment? It is not yet clear whether the defense agent and the vendor's autonomous components can cooperate, or whether they only act independently or interfere with one another. Our framework provides a testbed for these questions.

\subsubsection{Future work} Several downstream studies build on the framework rather than test it, since they concern the defense agent's behavior rather than the evaluation methodology. These are the cost-effectiveness of diverse proposer selections, richer attack-centric success metrics, a broader comparison of LLM backbones, hyperparameter sensitivity for agent optimality, and extension to other operator-deployed policies and autonomous-component tuning. The framework's closed-loop design and per-policy telemetry make each tractable.

\section{Conclusion}
We presented an evaluation framework for autonomous defense agents that harden a network by configuring a commercial EDR. It tests an agent in a closed loop, running real pentests and measuring each policy's effect at the finest granularity each EDR exposes. The framework separates the evaluation machinery from the agent under test through a pluggable proposer interface, and it enforces deployment verification and per-event policy attribution independently of the agent's choices. Between runs, it resets the lab machines and the EDR's cloud-side containment when present.

We instantiated the framework on the GOAD Active Directory lab, with NodeZero as the pentester and Microsoft Defender~XDR as the primary EDR, and benchmarked two LLM proposers, Claude Sonnet 4.6 and Cisco Foundation-Sec-8B. The campaign ran 20 pentests across four paired comparisons over a nine-day window, all under the same NodeZero attack configuration. We then extended the framework to CrowdStrike Falcon and SentinelOne Singularity to test whether the findings generalize. Three lessons emerged that help close the sim-to-real gap for defense-agent evaluation: (i) commercial EDR telemetry is built for SOC operator workflows, not for measuring policy effectiveness; (ii) per-policy attribution is needed to separate the defense agent's actions from the EDR's autonomous ones; and (iii) the EDR's autonomous behavior, such as Microsoft Defender~XDR's automatic attack disruption, varies across runs.

The deeper implication is a shift in how we frame the evaluation of defensive agents. Pre-AI evaluation methodology treats the defender's hardening tool as a passive, operator-defined surface and frames the agent's role as policy tuning atop it. In commercial EDR with autonomous AI, the tool is no longer passive. The EDR's autonomous layer (e.g., Microsoft Defender~XDR's automatic attack disruption) makes its own decisions and can produce much of the observed defensive output. Closing the sim-to-real gap for modern enterprise defense requires a deeper understanding of multi-agent defense dynamics. It also requires an evaluation methodology that treats the EDR as one of several autonomous components in the environment.

\section*{Acknowledgment}
The authors thank Cameron Denton for framework feedback, connector implementation, and GOAD lab setup, and Anton Foltz for debugging and technical expertise. We also thank Joshua Knox, Amy Villase\~{n}or, Naveen Sunkavally, and Snehal Antani for support and guidance, and Kainen Buch and Kidron Filbrun for tooling and cyber range support.

\section{LLM Usage Statement} \label{sec:llm-usage}
For reproducibility, deterministic Python scripts computed aggregated findings reported in this paper. The post-experiment per-event correlation between NodeZero attack telemetry and Microsoft Defender~XDR detection telemetry was inspected manually with assistance from Anthropic Claude Opus 4.7 (1M context). The inspection identified and corrected several enrichment-pipeline mis-attributions but did not alter the underlying raw-telemetry counts on which our findings are based. Anthropic Claude Opus 4.7 (1M context) via Claude Code was also used to aid in code implementation. LLMs were used for editorial purposes in this manuscript, and all outputs were inspected by the authors to ensure accuracy and originality.

\bibliographystyle{IEEEtran}
\bibliography{references}

\appendices

\section{Glossary}
\label{app:glossary}
This glossary collects the EDR and evaluation terms used throughout the paper for readers outside the endpoint-security domain. Table~\ref{tab:glossary} defines each term.

\begin{table*}[b!]
\centering
\caption{Glossary of EDR and evaluation terms.}
\label{tab:glossary}
\footnotesize
\setlength{\tabcolsep}{6pt}
\begin{tabularx}{\textwidth}{@{}l X@{}}
\toprule
Term & Definition \\
\midrule
EDR / XDR & Endpoint (Extended) Detection and Response: a host sensor that logs activity, raises alerts, and can block; XDR adds identity, email, and cloud signals. \\
MDE & Microsoft Defender for Endpoint: the endpoint component of Microsoft Defender XDR. \\
ASR & Attack Surface Reduction: Microsoft Defender rules that block classes of risky behavior, such as LSASS credential theft and PSExec/WMI execution. \\
PSExec & A Sysinternals tool for remote command execution over SMB; targeted, with WMI, by one ASR rule. \\
WMI & Windows Management Instrumentation: a Windows remote-management interface used for command execution; paired with PSExec in the ASR rule. \\
AIR & Automated Investigation and Response: Defender's autonomous investigation-and-remediation layer, gated by an automation level (full by default); distinct from attack disruption, which performs the containment. \\
Attack Disruption & Automatic Attack Disruption: Defender's built-in, incident-level containment that isolates hosts and contains user accounts (the \texttt{ContainedUser*} events), independent of the AIR automation level. \\
SOC & Security Operations Center: the analysts who triage EDR alerts. \\
MITRE ATT\&CK & A public catalog of attacker tactics and techniques. \\
APT & Advanced Persistent Threat: a sophisticated, sustained adversary. \\
GOAD & Game of Active Directory: a deliberately vulnerable Active Directory lab. \\
Domain controller & The server that holds Active Directory accounts and handles logins for its domain. \\
DA & Domain admin: a domain-administrator account or credential; a DA credential grants control of the domain. \\
ADCS & Active Directory Certificate Services: Microsoft's service for issuing certificates, often misconfigured in ways that allow privilege escalation. \\
SMB & Server Message Block: the Windows file- and printer-sharing protocol. \\
MSSQL & Microsoft SQL Server: Microsoft's database server. \\
IIS & Internet Information Services: Microsoft's web server. \\
NodeZero & Horizon3's autonomous penetration-testing product; the attacker in our experiments. \\
LSASS & The Windows process that holds credentials in memory, a common credential-theft target; LSA protection guards it. \\
LSA & Local Security Authority: the Windows subsystem that enforces security policy; LSA protection guards LSASS. \\
WDAC / PUA & Windows Defender Application Control (application allow/deny) and Potentially Unwanted Application protection. \\
Proposer & The wrapper around the defense agent that applies its chosen policies to the EDR each iteration. \\
LLM & Large Language Model: the model backbone driving a proposer (Claude Sonnet 4.6 and Cisco Foundation-Sec-8B in our benchmark). \\
Pentest & One NodeZero penetration-test run. \\
Closed loop & The repeated cycle of attack, measure, propose, and deploy that drives each experiment. \\
Storyline & SentinelOne's process-lineage graph used to attribute events; CrowdStrike's equivalent uses process-tree lineage. \\
Network Containment / Quarantine & CrowdStrike / SentinelOne persistent host-isolation states that hold a host offline until released. \\
\bottomrule
\end{tabularx}
\end{table*}

\section{Reproducibility and Commercial Dependencies}\label{app:repro} Our study relies on commercial products, which affects both its cost and its reproducibility. The lab is GOAD v3~\cite{goad} (commit \texttt{1bd35d9}) on eight Azure VMs, each \texttt{Standard\_B2s} except the Exchange member server (\texttt{Standard\_B4ms}) and the Linux host (\texttt{Standard\_B1s}); each EDR was tested on its own identically-configured GOAD instance. The pentester and all three EDRs require paid subscriptions: NodeZero (Horizon3, Elite tier), Microsoft Defender XDR (Microsoft~365 E5), CrowdStrike Falcon (Enterprise), and SentinelOne Singularity Complete. The two defense-agent proposers are Claude Sonnet 4.6, accessed through the Anthropic Messages API, and Cisco Foundation-Sec-8B, an open-weight security model served locally via Ollama as a Q8 GGUF through an OpenAI-compatible endpoint; their decoding settings and budget are in Table~\ref{tab:hyperparams}. The framework extends to other pentesters and EDRs, open-source or commercial, via a connector that emits telemetry in the standard format the defense agent consumes, though correlation, which maps a given attacker's actions to a given EDR's telemetry, must be built for each pairing of the two.

\begin{table*}[t]
  \centering
  \begin{minipage}[t]{0.48\textwidth}
  \renewcommand{\arraystretch}{1.1}
  \centering
  \caption{CrowdStrike Falcon prevention-policy action space.}
  \label{tab:cs-catalog}
  \footnotesize
  \begin{tabular}{lcp{3.0cm}}
  \toprule
  Setting group & Ct. & Modes \\
  \midrule
  ML detection/prevention sliders & 11 & 5 levels (Disabled to Extra Aggressive) \\
  Behavioral prevention toggles   & 56 & on / off \\
  \bottomrule
  \end{tabular}
  \end{minipage}\hfill
  \begin{minipage}[t]{0.48\textwidth}
  \renewcommand{\arraystretch}{1.1}
  \centering
  \caption{SentinelOne Singularity policy action space.}
  \label{tab:s1-catalog}
  \footnotesize
  \setlength{\tabcolsep}{4pt}
  \begin{tabular}{@{}p{2.6cm}c p{3.4cm}@{}}
  \toprule
  Setting group & Ct. & Modes \\
  \midrule
  Mitigation mode (standard, suspicious) & 2  & detect / protect \\
  Auto-mitigation action                 & 1  & none / kill / quarantine / rollback / remediate \\
  Detection engines                      & 14 & on / off \\
  Behavioral toggles                     & 25 & on / off \\
  Snapshot                               & 1  & on / off \\
  \bottomrule
  \end{tabular}
  \end{minipage}
  \end{table*}

\section{Per-Vendor Policy Action Spaces}
\label{app:action-spaces}

Section~\ref{sec:methods} describes the Microsoft Defender XDR policy catalog (Table~\ref{tab:catalog}) that defines the agent's action space. Tables~\ref{tab:cs-catalog} and~\ref{tab:s1-catalog} summarize the same catalog for CrowdStrike Falcon and SentinelOne Singularity. These catalogs let the agent run the closed loop on each vendor. On CrowdStrike, the agent toggles settings inside the prevention policy: 11 machine-learning sliders, each with five sensitivity levels, and 56 behavioral toggles. On SentinelOne, the agent sets the mitigation mode (detect or protect) for standard and suspicious threats. It also configures the auto-mitigation action, 14 detection engines, 25 boolean toggles, and snapshots. As on Microsoft Defender XDR, the agent deploys only the subset whose techniques overlap the weaknesses NodeZero exercised: the \texttt{CredentialDumping} toggle on CrowdStrike and the \texttt{mitigation\_mode} policy on SentinelOne.

\section{Per-Vendor Closed-Loop Results}
\label{app:vendor-results}

Tables~\ref{tab:cs-results}, \ref{tab:s1-results}, and~\ref{tab:defender-results} give the full per-run results for all three EDR vendors; Section~\ref{sec:results} interprets them. Each table lists, for every pentest in the closed loop, the policy the agent deployed that round, the number of NodeZero weaknesses found, the total EDR events, the share of actions with correlated telemetry, and the number of actions the EDR blocked or detected. Table~\ref{tab:cs-results} and Table~\ref{tab:s1-results} report the CrowdStrike Falcon and SentinelOne Singularity runs (three and five pentests, both with the Claude Sonnet 4.6 proposer). Table~\ref{tab:defender-results} reports the full Microsoft Defender XDR campaign across all four paired trials and both proposers.

\begin{table*}[!b]
\centering
\caption{CrowdStrike Falcon closed-loop results (Claude Sonnet 4.6 proposer).}
\label{tab:cs-results}
\small
\setlength{\tabcolsep}{5pt}
\begin{tabular}{@{}lllrrrrr@{}}
\toprule
Run & Proposer & Recommended Policy & N0 Weak. & Events & Corr. Tel. \% & Kills & Handle Downgr. \\
\midrule
1 (baseline) & --- & --- & 126 & 163 & 16.6 & 0 & 0 \\
2 & Sonnet 4.6 & CredDump (block) & 133 & 223 & 24.2 & 2 & 1 \\
3 (converg.) & Sonnet 4.6 & CredDump (block) & 136 & 223 & 24.2 & 1 & 2 \\
\bottomrule
\end{tabular}
\end{table*}

\begin{table*}[!b]
\centering
\caption{SentinelOne Singularity closed-loop results (Claude Sonnet 4.6 proposer).}
\label{tab:s1-results}
\small
\setlength{\tabcolsep}{5pt}
\begin{tabular}{@{}lllrrrr@{}}
\toprule
Run & Proposer & Recommended Policy & N0 Weak. & Events & Corr. Tel. \% & Mitigated \\
\midrule
1 (baseline) & --- & --- & 118 & 183 & 12.0 & 0 \\
2 & Sonnet 4.6 & MitigationMode (detect) & 123 & 180 & 10.0 & 0 \\
3 & Sonnet 4.6 & no new pick & 126 & 174 & 12.6 & 0 \\
4 & Sonnet 4.6 & MitigationMode (protect) & 105 & 93 & 21.5 & 24 \\
5 & Sonnet 4.6 & MitigationMode (protect) & 94 & 53 & 18.9 & 11 \\
\bottomrule
\end{tabular}
\end{table*}

\begin{table*}[!tb]
\centering
\caption{Microsoft Defender XDR closed-loop results, all iterations.}
\label{tab:defender-results}
\small
\setlength{\tabcolsep}{5pt}
\begin{tabular}{@{}lllrrrr@{}}
\toprule
Run & Proposer & Recommended Policy & N0 Weak. & Events & Corr. Tel. \% & Blocked Act. \\
\midrule
P1 baseline & --- & --- & 96 & 1790 & 98.9 & 46 \\
\cmidrule(lr){1-7}
P1 iteration 1 & Sonnet & LSASS+LSA block & 83 & 1500 & 100.0 & 64 \\
P1 iteration 2 & Sonnet & PSExec block & 81 & 57 & 0.0 & 0 \\
P1 iteration 3 (final) & Sonnet & LSASS/PSExec log, LSA off & 80 & 1033 & 100.0 & 8 \\
\cmidrule(lr){1-7}
P1 iteration 1 & F-Sec & LSA off, PSExec block & 73 & 560 & 86.1 & 6 \\
P1 iteration 2 (final) & F-Sec & LSASS+LSA block & 78 & 560 & 86.1 & 8 \\
\addlinespace[2pt]
\midrule[\heavyrulewidth]
\addlinespace[2pt]
P2 baseline & --- & --- & 95 & 1475 & 100.0 & 0 \\
\cmidrule(lr){1-7}
P2 iteration 1 & Sonnet & LSASS+LSA block & 79 & 1205 & 100.0 & 0 \\
P2 iteration 2 (final) & Sonnet & PSExec block & 81 & 1560 & 89.7 & 0 \\
\cmidrule(lr){1-7}
P2 iteration 1 & F-Sec & LSA off, PSExec log & 83 & 1120 & 100.0 & 4 \\
P2 iteration 2 & F-Sec & LSASS log, PSExec block & 119 & 3078 & 99.9 & 0 \\
P2 iteration 3 (final) & F-Sec & LSASS+LSA block & 95 & 2282 & 100.0 & 0 \\
\addlinespace[2pt]
\midrule[\heavyrulewidth]
\addlinespace[2pt]
P3 baseline & --- & --- & 126 & 3479 & 96.0 & 0 \\
\cmidrule(lr){1-7}
P3 iteration 1 (final) & Sonnet & LSASS+LSA block & 116 & 4260 & 97.7 & 0 \\
\cmidrule(lr){1-7}
P3 iteration 1 & F-Sec & LSA off, PSExec block & 123 & 4020 & 97.0 & 0 \\
P3 iteration 2 (final) & F-Sec & LSASS+LSA block & 118 & 3502 & 95.9 & 0 \\
\addlinespace[2pt]
\midrule[\heavyrulewidth]
\addlinespace[2pt]
P4 baseline & --- & --- & 123 & 3136 & 95.7 & 0 \\
\cmidrule(lr){1-7}
P4 iteration 1 & Sonnet & LSASS+LSA block & 91 & 1960 & 100.0 & 0 \\
P4 iteration 2 (final) & Sonnet & PSExec block & 87 & 1780 & 97.8 & 0 \\
\cmidrule(lr){1-7}
P4 iteration 1 (final) & F-Sec & LSA off & 127 & 3377 & 99.2 & 117 \\
\bottomrule
\end{tabular}
\end{table*}

\section{Per-Vendor Attack-Centric Outcomes}
\label{app:attack-outcomes}

From NodeZero's raw attack results, not the EDR's view, we report what it achieved against each configuration. A host is compromised when NodeZero proved credentials on it or bypassed its security control. We report compromised hosts (of 8), compromised domain controllers (of 3), compromised domains (of 3), stolen credentials (with proof or recovered material), domain-admin credentials, and ADCS privilege-escalation findings. Tables~\ref{tab:cs-attack}, \ref{tab:s1-attack}, and~\ref{tab:defender-attack} give the per-run results. Section~\ref{sec:results} interprets them: neither agent reduced compromise on CrowdStrike Falcon or SentinelOne Singularity, and Microsoft Defender XDR's per-run variation reflects its autonomous layer, not the deployed policy.

\begin{table}[h]
\centering
\caption{CrowdStrike Falcon: attack-centric outcomes (raw NodeZero results). Footprint constant across policies.}
\label{tab:cs-attack}
\footnotesize
\setlength{\tabcolsep}{3pt}
\begin{tabular}{@{}ll rrr rrr@{}}
\toprule
Run & Proposer & Hosts & DCs & Dom. & Creds & DA & PrivEsc \\
\midrule
baseline (detect-only) & ---    & 7 & 3 & 3 & 109 & 15 & 2 \\
CredDump (block)       & Sonnet & 7 & 3 & 3 & 158 & 14 & 2 \\
convergence (block)    & Sonnet & 7 & 3 & 3 & 171 & 15 & 3 \\
\bottomrule
\end{tabular}
\end{table}

\begin{table}[h]
\centering
\caption{SentinelOne Singularity: attack-centric outcomes (raw NodeZero results). Footprint constant; protect mode trims raw creds only.}
\label{tab:s1-attack}
\footnotesize
\setlength{\tabcolsep}{3pt}
\begin{tabular}{@{}ll rrr rrr@{}}
\toprule
Run & Proposer & Hosts & DCs & Dom. & Creds & DA & PrivEsc \\
\midrule
baseline (detect)     & ---    & 7 & 3 & 3 & 118 & 15 & 3 \\
detect                & Sonnet & 7 & 3 & 3 & 118 & 16 & 3 \\
detect (no new pick)  & Sonnet & 7 & 3 & 3 & 105 & 15 & 2 \\
protect               & Sonnet & 7 & 3 & 3 & 79  & 14 & 2 \\
protect (convergence) & Sonnet & 7 & 3 & 3 & 71  & 13 & 2 \\
\bottomrule
\end{tabular}
\end{table}

\section{Silent-Bypass Rates}
\label{app:silent-bypass}
Table~\ref{tab:silent-bypass} gives the per-run silent-bypass rate under two definitions of an EDR-evaded action. Under block-only evasion, an EDR-evaded action is one the EDR did not block. Under detection evasion, an EDR-evaded action is one the EDR neither blocked nor raised a named detection on; raw telemetry never counts under either. All rates are computed over admissible actions: NodeZero actions the EDR could observe given the delivery method and host role. Actions with no EDR expectation, such as network probes from the attacker host, are excluded; they are unblocked by definition, so the reported rates are a conservative measure of attacker stealth. Table~\ref{tab:silent-bypass} lists both rates per run. Section~\ref{sec:results} discusses what the two definitions imply and why Microsoft Defender XDR is measured on the automatic attack disruption runs.

The detection count is per-action for all three vendors: we tie each named detection to the NodeZero action it describes. CrowdStrike Falcon and Microsoft Defender XDR also record blocks we can tie to a single action. For Microsoft Defender XDR we mark an action as blocked when a named block lands within $\pm$300 seconds of it on the same host. We measure it only on the attack disruption baseline runs, where the containment event is both the block and the only named detection, so its two rates are equal. CrowdStrike Falcon records block status within each detection, and we attribute each enforcement to the action its detection describes.

SentinelOne Singularity is the exception. Its mitigations are threat-level and do not map to single actions, so we cannot label individual actions as blocked. Instead we use the run total, subtracting the enforced process kills (24 in protect run~4, 11 in protect run~5) from the admissible actions. One mitigation can span several actions, so this total is a lower bound on blocked actions, which makes the SentinelOne Singularity block-only rates upper bounds. The detect runs have zero kills, so their block-only evasion is 100\%.

Microsoft Defender XDR no-reset run~3 reaches 100\% block-only for the opposite reason: no admissible action matched a named block. Table~\ref{tab:air-reset} still records 839 \texttt{ContainedUser*} events for that run, but these are repeated blocks of accounts that remain contained from earlier state, and none map to a distinct admissible action.

\section{Automatic Attack Disruption State Carryover}
\label{app:air-carryover}

Table~\ref{tab:air-reset} gives the full per-run counts for the reset and no-reset automatic attack disruption baseline experiment referenced in Section~\ref{subsec:air-variation}. The start-state columns measure carried-over containment; the new-containment columns measure per-run automatic attack disruption activity.

\begin{table*}[t]
\centering
\begin{minipage}[t]{0.5\textwidth}
\centering
\caption{Microsoft Defender XDR: attack-centric outcomes (raw NodeZero results), all 20 pentests. Footprint flips between 7/3/3 and 4/2/2 independent of policy, reflecting the autonomous attack disruption layer rather than the agent (Section~\ref{subsec:air-variation}).}
\label{tab:defender-attack}
\footnotesize
\setlength{\tabcolsep}{3pt}
\resizebox{\linewidth}{!}{%
\begin{tabular}{@{}ll rrr rrr@{}}
\toprule
Run & Proposer & Hosts & DCs & Dom. & Creds & DA & PrivEsc \\
\midrule
Pair 1 baseline     & ---    & 7 & 3 & 3 & 123 & 2  & 4 \\
\cmidrule(lr){1-8}
iteration 1         & Sonnet & 4 & 2 & 2 & 99  & 0  & 4 \\
iteration 2         & Sonnet & 4 & 2 & 2 & 103 & 0  & 3 \\
iteration 3 (final) & Sonnet & 4 & 2 & 2 & 85  & 0  & 4 \\
\cmidrule(lr){1-8}
iteration 1         & F-Sec  & 4 & 2 & 2 & 83  & 0  & 3 \\
iteration 2 (final) & F-Sec  & 4 & 2 & 2 & 89  & 0  & 3 \\
\midrule[\heavyrulewidth]
Pair 2 baseline     & ---    & 4 & 2 & 2 & 112 & 0  & 4 \\
\cmidrule(lr){1-8}
iteration 1         & Sonnet & 7 & 3 & 3 & 84  & 0  & 4 \\
iteration 2 (final) & Sonnet & 4 & 2 & 2 & 98  & 4  & 3 \\
\cmidrule(lr){1-8}
iteration 1         & F-Sec  & 4 & 2 & 2 & 90  & 0  & 4 \\
iteration 2         & F-Sec  & 7 & 3 & 3 & 227 & 10 & 3 \\
iteration 3 (final) & F-Sec  & 7 & 3 & 3 & 123 & 4  & 3 \\
\midrule[\heavyrulewidth]
Pair 3 baseline     & ---    & 7 & 3 & 3 & 247 & 10 & 4 \\
\cmidrule(lr){1-8}
iteration 1 (final) & Sonnet & 7 & 3 & 3 & 227 & 11 & 3 \\
\cmidrule(lr){1-8}
iteration 1         & F-Sec  & 7 & 3 & 3 & 243 & 13 & 3 \\
iteration 2 (final) & F-Sec  & 7 & 3 & 3 & 218 & 11 & 3 \\
\midrule[\heavyrulewidth]
Pair 4 baseline     & ---    & 7 & 3 & 3 & 248 & 11 & 3 \\
\cmidrule(lr){1-8}
iteration 1         & Sonnet & 4 & 2 & 2 & 113 & 7  & 4 \\
iteration 2 (final) & Sonnet & 4 & 2 & 2 & 109 & 5  & 3 \\
\cmidrule(lr){1-8}
iteration 1 (final) & F-Sec  & 7 & 3 & 3 & 247 & 14 & 4 \\
\bottomrule
\end{tabular}%
}
\end{minipage}\hfill
\begin{minipage}[t]{0.45\textwidth}
\centering
\caption{Silent-bypass rate per run, under two definitions of an EDR-evaded action. \emph{Block-only evasion}: an EDR-evaded action is one the EDR did not block. \emph{Detection evasion}: an EDR-evaded action is one the EDR neither blocked nor raised a named detection on. For Microsoft Defender XDR, an action is blocked when a named block lands within $\pm$300 seconds on the same host. For CrowdStrike Falcon, we read block status from each detection and tie it to that action. For SentinelOne Singularity, mitigations are threat-level, so we subtract the enforced-kill count from the admissible actions.}
\label{tab:silent-bypass}
\scriptsize
\setlength{\tabcolsep}{4pt}
\begin{tabular}{@{}l rr@{}}
\toprule
 & \multicolumn{2}{c}{Silent-bypass rate} \\
\cmidrule(lr){2-3}
Run & Block-only & Detection \\
\midrule
\multicolumn{3}{@{}l}{\textit{CrowdStrike Falcon}} \\
baseline (run 1)       & 100.0\% & 99.0\% \\
CredDump block (run 2) & 99.9\%  & 99.6\% \\
convergence (run 3)    & 99.9\%  & 99.6\% \\
\cmidrule(lr){1-3}
\multicolumn{3}{@{}l}{\textit{SentinelOne Singularity}} \\
detect (run 1)   & 100.0\% & 68.9\% \\
detect (run 2)   & 100.0\% & 67.9\% \\
detect (run 3)   & 100.0\% & 72.3\% \\
protect (run 4)  & 91.4\%  & 76.6\% \\
protect (run 5)  & 95.7\%  & 81.0\% \\
\cmidrule(lr){1-3}
\multicolumn{3}{@{}l}{\textit{Microsoft Defender XDR}} \\
run 1 (shared)   & 97.8\%  & 97.8\% \\
\addlinespace[3pt]
reset (run 2)    & 94.8\%  & 94.8\% \\
reset (run 3)    & 95.4\%  & 95.4\% \\
\addlinespace[3pt]
no-reset (run 2) & 94.6\%  & 94.6\% \\
no-reset (run 3) & 100.0\% & 100.0\% \\
\bottomrule
\end{tabular}
\end{minipage}
\end{table*}

\begin{table*}[!tb]
\centering
\caption{Automatic attack disruption state carryover. Repeated no-agent Microsoft Defender XDR baseline (identical no-policy config), run three times with state reset between runs and three times without, both forked from one shared clean run (run 1). Users and machines contained at run start measure carried-over state; new contains and isolations measure per-run activity.}
\label{tab:air-reset}
\small
\setlength{\tabcolsep}{4pt}
\begin{tabular}{@{}lrrrrrrrr@{}}
\toprule
Run & N0 Events & Detections & Telemetry & Users Cont.\ (start) & New Cont. & Mach.\ Iso.\ (start) & New Iso. & Contain Blocks \\
\midrule
\multicolumn{9}{@{}l}{\textit{Reset between runs}} \\
run 1 & 3711 & 7 & 68 & 0 & 6 & 0 & 0 & 121 \\
run 2 & 5339 & 22 & 62 & 0 & 4 & 0 & 1 & 60 \\
run 3 & 4709 & 20 & 67 & 0 & 3 & 0 & 0 & 106 \\
\cmidrule(lr){1-9}
\multicolumn{9}{@{}l}{\textit{No reset between runs}} \\
run 1 & 3711 & 7 & 68 & 0 & 6 & 0 & 0 & 121 \\
run 2 & 4261 & 17 & 83 & 6 & 6 & 0 & 1 & 36 \\
run 3 & 2926 & 0 & 44 & 12 & 7 & 1 & 0 & 839 \\
\bottomrule
\end{tabular}
\end{table*}

\end{document}